# Accessibilité des sites web pour les personnes présentant une incapacité visuelle
_______________________________________________________________________________________

Katerine ROMEO[1]       Edwige Pissaloux[2]       Frédéric Serin[3]

Katerine.Romeo@univ-rouen.fr   Edwige.Pissaloux@univ-rouen.fr   Frederic.Serin@univ-lehavre.fr

[1] IUT de Rouen, [2]Université de Rouen Normandie, [3]Université du Havre Normandie
Laboratoire LITIS/CNRS FR 3638

**Thèmes –** *Informatique - Robotique - Imagerie*

**Résumé –** *L'accessibilité des sites web pour les personnes présentant une incapacité visuelle (PPIV) est malmenée avec les lecteurs d'écran qui ne sont pas toujours adaptés à l'interactivité que les technologies web/multimédia actuelles nécessitent. Ce papier analyse les difficultés d'accès à l'information avec l'utilisation des différentes technologies et avec les recommandations existantes pour la conception des sites web accessibles à tous. Les résultats d'évaluation préliminaire avec les PPIV du site web de notre projet ACCESSPACE sont présentés.*

**Mots-Clés –** *E-accessibilité, information visuelle, information tactile, déficience visuelle.*

# 1 Introduction

Nous sommes actuellement confrontés à un illettrisme numérique d'un nombre grandissant de personnes présentant une incapacité visuelle (PPIV, seniors compris) car l'accessibilité des sites web et des documents numériques/multimédia ne suit pas les évolutions de technologies de conception de logiciels. Pourtant, des efforts sont faits depuis 2009, avec le Référentiel Général d'Accessibilité pour les Administrations (RGAA); avec les recommandations d'AccessiWeb et d'Opquast qui a établi une liste de 226 critères pour un site de qualité et accessible. Les recommandations ne sont pas toujours bien comprises ou pas correctement implantées par les concepteurs et réalisateurs de cites.

Or, selon l'OMS (Organisation Mondiale de la Santé) déjà 253 millions de personnes présentent une incapacité visuelle dans le monde [1] et ce nombre continue à croître avec la société numérique et avec l'allongement de la vie. Différents types de déficience doivent être considérés ; en effet, parmi les PPIV répertoriées 4,5% environ ont une déficience de perception de couleur, 4% de personnes ont une vision tubulaire ou un champ de vue réduit ou encore des problèmes d'acuité visuelle, et 0,6% sont des non-voyants. Tous, mais les PPIV en premier, ont des difficultés à localiser l'information sur les sites web et à comprendre le contenu indiqué. Ces difficultés peuvent être induites par des technologies d'assistance logicielles (comme des lecteurs d'écran), et matérielles (comme des tablettes braille).

Cette communication analyse les méthodes d'exploration d'un document (Section 2). Les faiblesses de technologies les plus populaires pour accéder au document sont rappelées dans la Section 3. Les résultats de l'évaluation préliminaire de ce dernier point avec les déficients visuels sont analysés dans la Section 4. La Section 5 conclut et indique quelques prolongements possibles du travail présenté.

# 2 Exploration d'un document numérique

L'exploration d'un document numérique est une fonction de deux variables : les stratégies cognitives de l'exploration du texte lu et de l'organisation spatiale du texte. La lecture naturelle met en œuvre les différentes stratégies cognitives pour comprendre un document avec une suite de processus cognitifs élémentaires [2]. Parmi ces processus on peut citer la saisie de l'information dans une phrase, l'intégration et la recherche de cohérence entre les phrases, la compréhension du texte en entier et l'identification des idées principales, l'élaboration du texte (qui s'appuie sur la prédiction, l'image mentale, la réponse affective) ; et aussi, la perception par le lecteur de ses capacités à comprendre le texte et à se réajuster s'il ne le comprend pas. Cependant, ces processus sont impactés par l'organisation de document : linéaire ou non.

La lecture d'un *document linéaire* (un livre par exemple) se fait de façon séquentielle. Les mots sont identifiés et la position de certains mots importants dans la page est mémorisée. Ce processus permet des retours en arrière à une place fixe et une bonne assimilation des informations. Le contexte restant stable, les processus d'appréhension et de mémorisation de l'information sont renforcés. Lorsqu'on lit un texte sur une page web, ce processus de mémorisation spatiale de mots est affaibli avec le déroulement vertical (scrolling), voire perdu avec les mises à jour instantanées de pages (travail collaboratif avec des documents partagés). Le lecteur est confronté à une instabilité spatiale et perd ses repères spatiaux. Consulter plusieurs pages en parallèle amène la confusion avec des répétitions et rend difficile la représentation mentale du contenu qui évolue en une *structure arborescente*.

Tous les éléments d'accès et d'exploration d'un document cités ci-dessus sollicitent grandement la vision et sont donc non exploitables par les PPIV privées de la perception visuo-spatiale. La navigation sur internet ou dans un document pour la PPIV se passe différemment : le lecteur d'écran parcourt séquentiellement un document. En effet, un lecteur de l'écran est une source unique de l'information (audio, TTS – texte-to-speech) ; comme il ne peut mettre en œuvre que la lecture séquentielle (mono-voix), il suppose que tout document a une structure linéaire et il provoque une perte de l'information sur l'organisation spatiale de celui-ci. Pour pallier à ces inconvénients certaines règles d'organisation spatiale ont été proposées. L'organisation par menus (déroulants) en est une. La navigation d'un site web à travers un menu placé tout en haut de la page d'accueil – à la racine de l'arborescence du document entier. Aussi, la recherche de l'information sera intimement liée à l'exploration de l'arborescence. La stratégie de son parcours est primordiale car elle conditionne la vitesse d'accès à l'information recherchée et la charge cognitive nécessaire durant cette recherche.

Force est de constater que les moteurs de parcours de l'arbre du document offrent surtout la stratégie de parcours systématique. En effet, la structure d'un document web type en HTML5 est une arborescence dont les sommets (nœuds) sont matérialisés avec des balises sémantiques ; ces balises devraient être situées à leur place de définition. *La figure 1* montre l'arborescence de la structure de la page web HTML5 sémantique.

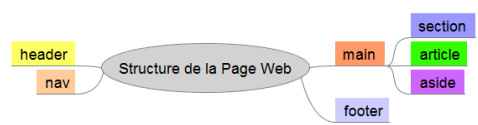

Figure 1 – Arborescence d'une page web HTML5 sémantique.

## 3 Technologies pour accéder au document.

Deux technologies sont les plus populaires auprès de PPIV pour accéder au contenu d'un document : un lecteur d'écran et une tablette à stimulation tactile.

Le lecteur d'écran JAWS (Job Access With Speech, 1989) accède au DOM (Document Object Model) et utilise les interfaces de programmation du navigateur pour avoir les informations sur les objets, leur rôle, leur nom et leur état afin de les annoncer au moment du parcours de la page [3]. Un autre lecteur d'écran pour Windows est NVDA (Non Visual Desktop Access) développé en 2007 en open source. Ces lecteurs d'écran, grâce aux balises HTML indiquent, tant bien que mal, l'organisation spatiale. Le contenu de la page est exploré séquentiellement en parcourant la liste de balises avec une synthèse vocale ou en braille.

Les tablettes à stimulation tactile affichent, le plus souvent, une information 2D sous l'une de formes à la fois braille, texture ou contours. Leur surface doit être explorée manuellement. Pour les contenus visuels comme des images ou des graphiques, il existe des prototypes de tablettes à retour tactile qui visent à produire des formes simples (approximées) d'informations visuelles [4]. Ces appareils de substitution visuo-tactiles utilisent différentes technologies, comme Stimtac de l'Université de Lille 1 ou la Hap2U, qui présentent une surface avec une vibration et un coefficient de friction plus ou moins intense lors de l'action de l'exploration (principe : action-perception)[5].

## 4 Evaluation préliminaire d'un site web

Nous avons réalisé un site web pour notre projet Accesspace (accesspace.univ-rouen.fr) en collaboration avec L'Espace Handicap de l'Université de Rouen. Sa première réalisation a été une occasion d'appréhender les difficultés rencontrées par des PPIV. La structure est linéaire avec un parcours systématique de l'arborescence attachée au site. Les règles RGAA ont été implantées. Quatre PPIV ayant différents degrés de perception visuelle (déficience tardive, vision tubulaire, vision périphérique, un aveugle de naissance) ont participé à l'évaluation de notre site avec 25 questions notées. Elles ont consulté le site avec un lecteur d'écran, et/ou sa transcription braille (Fig. 3). Pour les malvoyants, le site pouvait être agrandi sur un grand écran. Les participants donnaient une note sur 10 de différents points évalués.

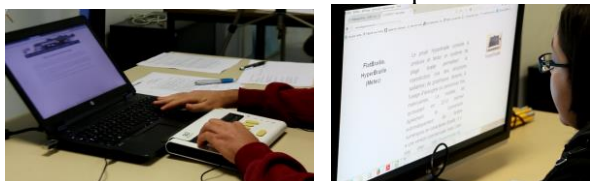

Figure 2 – Evaluation du site avec un lecteur d'écran et un clavier braille ou avec l'agrandissement du contenu.

L'évaluation a porté sur deux éléments : l'accessibilité à l'information et la compréhension du contenu du site. La possibilité de *navigation* sur le site (càd. la recherche des liens) avec le clavier ou la souris a été bien appréciée, ainsi que la tâche de retrouver *l'onglet* (la recherche des « zones cliquables ») qui intéressait le participant, en cas de confusion, a été considéré comme facile. L'accès au contenu d'images et de vidéo a été assez bien évalué mais les participants ont attendu leur meilleure présentation (par exemple avec une audiodescription). Par ailleurs, nous avons constaté que le contenu des pages n'était pas toujours compréhensible avec le lecteur d'écran, surtout quand la configuration de l'outil n'était pas optimisée et personnalisée.

## 5 Conclusion et perspectives

Cette communication a proposé une analyse de l'accessibilité aux informations sur le web et une méthode d'évaluation de l'accessibilité de site avec les PPIV. L'analyse de l'état de l'art montre que les systèmes TIC à parcours systématique de l'arborescence attachée à un document web ne permettent pas d'accéder à toute l'information, et que les normes d'accessibilité existantes ne considèrent pas toute la complexité d'un document.

Un standard de structuration logique du contenu devrait être guidé par sa représentation interne arborescente balisée, et tenir compte des mécanismes cognitifs de recherche des informations. L'évaluation préliminaire du site de notre projet ACCESSPACE confirme que le degré de déficience de chaque participant impacte les besoins en accessibilité ; aussi, la personnalisation d'un outil est fondamentale. Ces résultats devront être affinés avec un nombre plus important de participants qui devraient être repartis dans des groupes plus homogènes, notamment selon le degré de déficience afin de pouvoir utiliser les outils statistiques inférentiels et établir des règles d'accessibilité plus fiables. Finalement, des technologies multimodales d'accessibilité (tactile et audiodescription) devraient être développées.